\documentclass[onecolumn,a4paper,11pt]{cnops}
\usepackage{graphicx}
\usepackage{amsmath}

\usepackage[breaklinks]{hyperref}
\usepackage{natbib}
\usepackage{multirow}
\bibpunct{(}{)}{;}{a}{}{,}

\def\inst#1{$^{#1}$}

\newcommand{\mygi}{MyGIsFOS}

\newcommand{\kms}{$\rm km s ^{-1}$}
\newcommand*\farcs{\ensuremath{\overset{\prime\prime}{.}}}
\DeclareRobustCommand{\ion}[2]{\textup{#1\,\textsc{\lowercase{#2}}}}
\begin{document}

\title{Abundances of CNO in candidate young metal-poor stars}
%\author{Farina von Maintal\inst{1},
%\\
%\footnotesize
%Observatoire Gastronomique de Paris, 95 rue St. Honoré, 
%75001, Paris, France\\
%}
\noindent
\author{P.~Bonifacio\inst{1},
E.~Caffau\inst{1},
L.~Sbordone\inst{2},
L.~Monaco\inst{3},
L.~Lombardo\inst{4},
\\
R.~Lallement\inst{1},
M.~Spite\inst{1},
P.~Fran\c{c}ois\inst{5,6},
A.~Mucciarelli\inst{7}
\\
\noindent
\footnotesize
$^1$LIRA, Observatoire de Paris, Universit\'e PSL,Sorbonne Universit{\'e}, Universit{\'e} Paris Cit{\'e},\\
\footnotesize
 CY Cergy Paris Universit{\'e}, CNRS, 5 place Jules Janssen 92195 Meudon, France  \\
\footnotesize
$^2$European Southern Observatory, Alonso de Cordova 3109, Vitacura, Santiago, Chile\\
\footnotesize
$^3$ Universidad Andres Bello, Facultad de Ciencias Exactas, Departamento de F\'isica y Astronom\'ia\\
\footnotesize
 Instituto de Astrof\'isica, Autopista Concepci\'on-Talcahuano 7100, Talcahuano, Chile\\
\footnotesize
$^4$INAF-Osservatorio  Astronomico  di  Trieste,  Via  G.B.  Tiepolo  11,34143 Trieste, Italy\\
\footnotesize
$^5$LIRA, Observatoire de Paris, Universit{\'e} PSL, Sorbonne Universit{\'e}, Universit{\'e} Paris Cité,\\
\footnotesize
CY Cergy Paris Universit{\'e}, CNRS,75014 Paris, France\\
\footnotesize
$^6$UPJV, Universit\'e de Picardie Jules Verne, 33 rue St Leu, 80080 Amiens, France\\
\footnotesize
$^7$Dipartimento di Fisica e Astronomia, Universit\`a degli Studi di Bologna, Via Gobetti 93/2, I-40129 Bologna, Italy
\\
}
\maketitle

{\sl {\bf Keywords:} Stars: abundances -- Stars: Population II -- Galaxy: abundances  -- Galaxy: evolution}\hfill\\
~\\
{\bf Published:} \today \\

\begin{abstract}
  In this contribution we investigate the CNO abundances in five apparently young evolved stars,
  with the aim of discriminating between truly young stars and stars that were rejuvenated 
  by accreting mass from another star. Stars that have accreted mass are expected to show low C and O and a
  very low [C/O] ratio, as displayed by some  stars in the Globular Cluster 47 Tuc, that are believed
  to have undergone mass-transfer.
  In our sample the low [C/O] ratios observed appear to be compatible with their evolutionary
  status. There is thus no indication for these stars having accreted mass. 
\end{abstract}

\section{Introduction}

In the last years several metal-poor stars that are apparently young have 
been discovered \citep[seee e.g.][]{caffau2020,caffau2024,topos6}. Some of these 
could be young stars that were formed as a consequence of a starburst 
induced in  a gas rich  dwarf galaxy infalling into the Milky Way, 
as suggested by \citet{yang2024}. Others could be stars that have 
been rejuvenated by accreting mass from one or several other stars.
A possible diagnostic to discriminate the two cases is the [C/O]
ratio. 

\section{Target selection and observations}

In the Gaia\,DR3, we selected a sample of bright stars ($7\le G\le 13$) with transverse velocity in excess of 500 \kms,
parallax larger than three times the parallax error.
%and $7\le G\le 13$.
Among these stars we noticed a large number of stars, that, in the colour-magnitude
diagram, occupy the region where young evolved stars can be found \citep[see][figures 1 and 2]{bonifacio2024}.
%Our selected targets are shown in Fig.\,1, the left panel shows the colour magnitude diagram
%and the right panel the angular momentum versus the square root of the radial action.
%As comparison we show a set of very young metal-poor isochrones and a set of old isochrones. 

%\begin{figure}%{r}{0.5\textwidth} 
%\centering
%\includegraphics[width=8cm]{cmd_targets.png}
%\includegraphics[width=8cm]{Lz_Jr.png}
%
%\caption{In the left panel the colour-magnitude diagram for our targets, with a set
%of Parsec isochrones (Bressan et al. 2012) of metallicity --1.35. In the right panel the
%plot of angular momentum versus the square root of radial action. The units are kpc\,km/s for the x-axis
%and the same unit to the power 0.5 for the y-axis. Some of the targets can be associated
%to the Gaia-Sausage-Enceladus structure and some to the Sequoia structure.}
%\end{figure}

A subset of these stars was observed with UVES at the ESO 8.2\,m
telescope with the DIC2 437+760 setting, that covers the 
wavelength ranges  373–-499\,nm and 565–-946\,nm using a 
0\farcs{5} slit that provides a resolving power
of about 75\,000 in the blue and 70\,000 in the red arm.

In our contribution to this conference we shall discuss 
the CNO abundances of five apparently young stars whose
characteristics are reported in Table \ref{tab:targets}.
\begin{table}
\caption{Parameters of the program stars}
\label{tab:targets}
\begin{tabular}{rlrlllcl}
\hline
  \multicolumn{1}{c}{Gaia DR3 ID} &
  \multicolumn{1}{c}{NAME} &
  \multicolumn{1}{c}{Teff} &
  \multicolumn{1}{c}{log g} &
  \multicolumn{1}{c}{VTURB} &
  \multicolumn{1}{c}{[Fe/H]} &
  \multicolumn{1}{c}{S/N@420\,nm} &
  \multicolumn{1}{c}{Vrad} \\
  \multicolumn{1}{c}{} &
  \multicolumn{1}{c}{} &
  \multicolumn{1}{c}{K} &
  \multicolumn{1}{c}{log(cgs)} &
  \multicolumn{1}{c}{\kms} &
  \multicolumn{1}{c}{dex} &
  \multicolumn{1}{c}{~} &
  \multicolumn{1}{l}{\kms} \\
\hline
  3151624328675371264 & YMP015 & 4910 & 2.38 & 1.48 & --0.67 &  46 & \phantom{--0}26.2\\
  2716352208089611264 & YMP033 & 4600 & 1.00 & 2.06 & --1.67 & 70  &  --260.1 \\
  6897294199759073536 & YMP053 & 5229 & 2.43 & 1.68 & --1.47 & 72  & \phantom{1}--70.3 \\
  2796582811359131392 & YMP054 & 5151 & 2.10 & 1.84 & --2.00 & 94  & --157.2 \\
  722021080910867584 & YMP067 & 4546  & 0.65 & 2.26 & --2.41 & 80 & \phantom{1}--13.0  \\
  \hline\end{tabular}
\end{table}

\subsection{Stellar parameters and chemical abundances}

We used the Gaia photometry, corrected for reddening  according
to the maps of \citet{vergely2022} and extinction coefficients from the
KOALA database \citep{KOALA} to determine effective temperatures.
The photogeometric distances of 
\citet{BailerJones2021} were used with the Stefan-Boltzmann equation to determine
surface gravities. The whole procedure was iterative as described in detail by \citet{lombardo2021}, with
one difference: at each step we estimated ages and masses using the Bayesian inference code SPInS \citep{LebretonReese2020,spins2020}
and the assumed mass was updated. Only three iterations were necessary to converge, with variations of effective temperatures
being less than 10\,K and in surface gravity of less than 0.05\,dex.
To run SPInS we used the BaSTI evolutionary tracks enhanced in $\alpha$ elements by 0.4\,dex \citep{Pietrinferni2021}.
We adopted a flat prior of ages between 0 and 13.8\,Gyr 
to avoid ages larger than the age of the Universe.

At each iteration we run \mygi\ \citep{sbordone2014} to determine the chemical composition
of each star. The oxygen abundances were determined by \mygi\ using at least one
[OI] line (630.0\,nm, 636.3\,nm)
and at least one permitted \ion{O}{i} line
of the IR triplet (777.1\,nm, 777.3\,nm,777.5\,nm).
Unsuprisingly it was for the warmest star of the sample YMP054 that we kept
only two lines: the [OI] 630.0\,nm with an equivalent width of 0.714\,pm and the
\ion{O}{i} 777.1\,nm line with an equivalent width of 1.342\,pm.
For the other four stars we have either five or four \ion{O}{i} lines.
When there were NLTE corrections for the triplet lines by \citet{sitnova2013}
we applied them to the LTE abundances derived by \mygi. If there were no corrections
available at parameters near those of the stars we assumed a zero NLTE correction.
The largest root mean square deviation among the \ion{O}{i} lines is found
for YMP067, 0.12\,dex, over five lines, and for this star we did not apply any NLTE corrections.

For carbon \mygi\ determined the abundance from one or more of the IR \ion{C}{i} lines
at  833.5149\,nm, 872.7126\,nm, 906.1432 and 907.8278\,nm. 
The \ion{C}{i} lines used were: for YMP015 the lines at 833.5 and 872.7\,nm; for YMP033 the line at 906.1\,nm; for YMP053 the lines at 872.7 and 907.8\,nm; for YMP054 the lines at 833.5 and 906.1\,nm; for YMP067 no line.
%with the exception of YMP033 for which no atomic carbon line is detected. 
For all stars we determined the C abundance also from a $\chi^2$ fit to the CH $A^2\Delta-X^2\Pi$ molecular band, hereafter the  G-band \citep{Fraunhofer}.
For all stars, except YMP054, the C abundance from the G-band is significantly 
lower that that determined from the atomic lines. 
We do not have at our disposal NLTE computations for the IR carbon
lines that cover the stellar parameters of the present sample. 
However computations of
\citet[][]{Alexeeva2014} suggest there
should not be 
a large NLTE effect, of the order of --0.1\,dex.
\cite{review} pointed out the case of three extremely-metal poor
stars, G\,64-12, G\,64-37 and G\,275-4, for which a disturbing
discrepancy exists between the carbon abundances derived from the 
atomic lines, even with the most sophisticated 3D NLTE modelling
and the G-band. We are here in a much higher metallicity regime, but the underlying
reason could be the same. A suspected reason are 3D NLTE effects on
one or several of the diatomic molecules that are linked to CH through the
chemical network: CH, CO, NH, OH, CN.

Considering the wavelength range available to us the only
possibility to determine the nitrogen abundancce
is through the CN $B^2\Sigma^+-X^2\Sigma^+(0-0)$ band at 388.4\,nm
and through the CN $B^2\Sigma^+-X^2\Sigma^+(0-1)$ band at 421.6\,nm.
The difficulty in modelling the 388.3\,nm band, affected both by 3D and NLTE effects
\citep[][]{Bonifacio2013,Mount1975} is well known, and because of lack
of investigations, we do not know if the 421.6\,nm band is any better.
In our sample the 421.6 nm band is detectable only in YMP015  and YMP033,
in all cases there is a considerable discrepancy. Furthermore
for the stars for which we had a discrepancy in the C abundances from atomic line and
G-band we computed the N abundance for both possible C abundances. 
We note that all our fits to the CN 388\,nm  are poor, we are able to reproduce the wings of
the band, but not the core. If we force the fit to the core, we lose the wings.
This same problem was found by \citet{spite2005}, suggesting that our model atmospheres 
and line transfer computations are not adequate to fully describe this band.
The usual suspects are a chromospheric emission in the core and 3D-NLTE effects.
Our results for the CNO elements are summarised in Table\,\ref{tab:cno}

\begin{table}[]
\setlength{\tabcolsep}{2pt}
    \caption{CNO abundances for our program stars}
    \centering
    \begin{tabular}{lllllllll}
\hline
Star  & A(O)  & $\sigma[A(0)]$ & $N_O$ & A(C)$_{CI}$ & A(C)$_{CH}$ & A(N)$_{388\_CI}$ & A(N)$_{388\_CH}$  & A(N)$_{421\_CH}$ \\ 
\hline
YMP015 & 8.36 & 0.03           & 5     & 7.81        & 7.70      &                  & 7.30             & 7.60             \\
YMP033 & 7.64 & 0.06           & 4     &  6.49       & 6.28      &                  & 6.75             &                  \\ 
YMP053 & 7.91 & 0.04           & 4     & 7.21        & 6.73      & 7.20             & 6.71             &                   \\ 
YMP054 & 7.31 & 0.004          & 2     & 6.61        & 6.57      & 6.22$^a$         & 6.22$^a$         &                    \\ 
YMP067 & 7.10 & 0.12           & 5     &             & 5.54      &                  & 6.37             &                    \\ 
\hline
\multicolumn{9}{l}{We assumed A(C) 6.52, that is the average of the abundance from \ion{C}{i} and CH.}
    \end{tabular}
    \label{tab:cno}
\end{table}

\section{Discussion and conclusions}
%\begin{figure}
%    \centering
%   \resizebox{0.8\textwidth}{!}{\includegraphics{age_mass_sesto.pdf}}
%    \caption{Mass vs. age diagram for our program stars}
%    \label{fig:agemass}
%\end{figure}
\begin{figure}
    \centering
   \resizebox{0.8\textwidth}{!}{\includegraphics{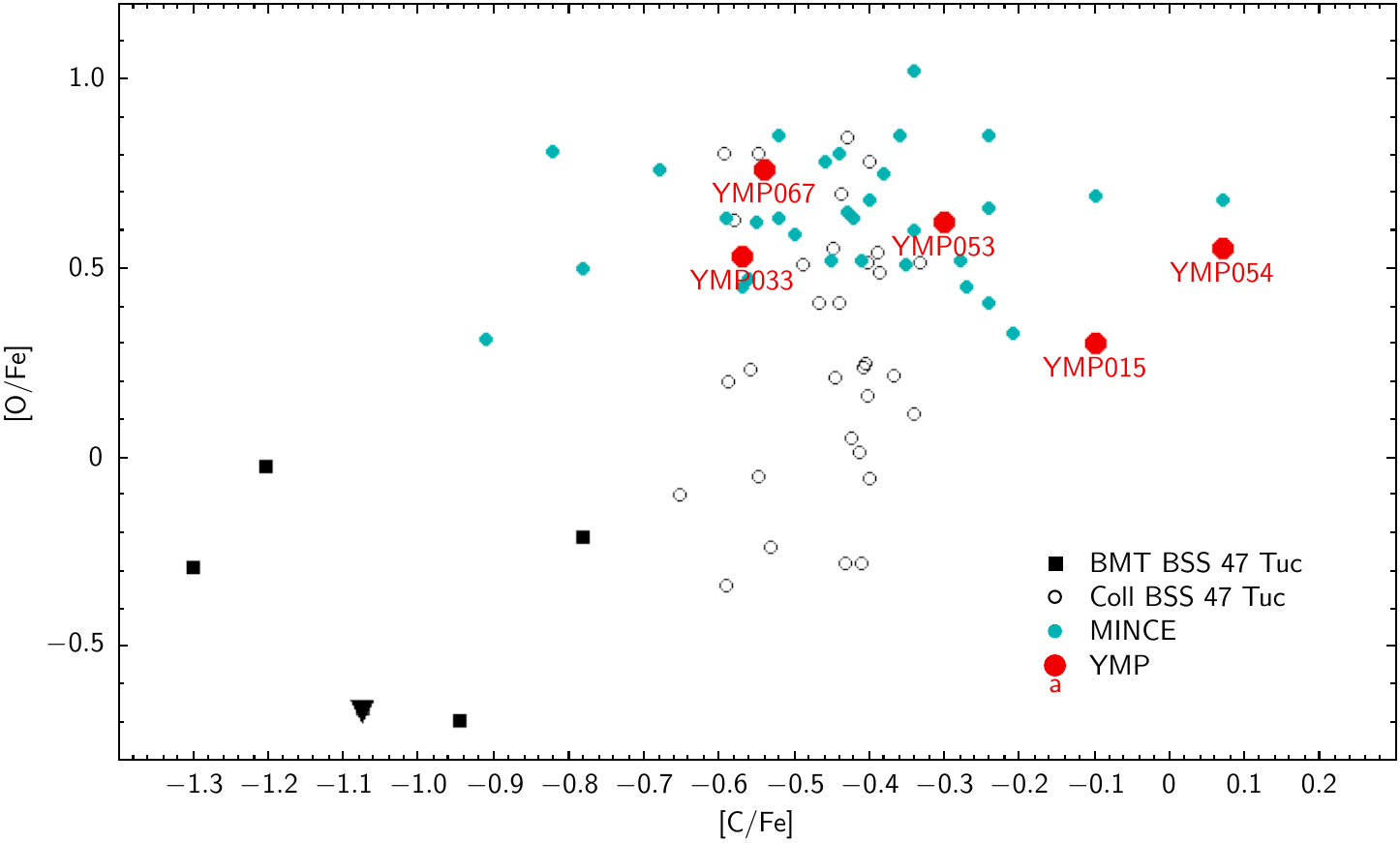}}
    \caption{ The [O/Fe] vs. [C/Fe] diagram for our stars. As reference stars we use
    the BSS stars in 47\,Tuc from \citep[][black, open symbols collisional BSS, filled symbols mass transfer BSS]{ferraro2006}
    and from the MINCE collaboration (Lombardo et al., these proceedings, cyan symbols).} 
    \label{fig:cfe_ofe}
\end{figure}
\begin{figure}
    \centering
   \resizebox{0.8\textwidth}{!}{\includegraphics{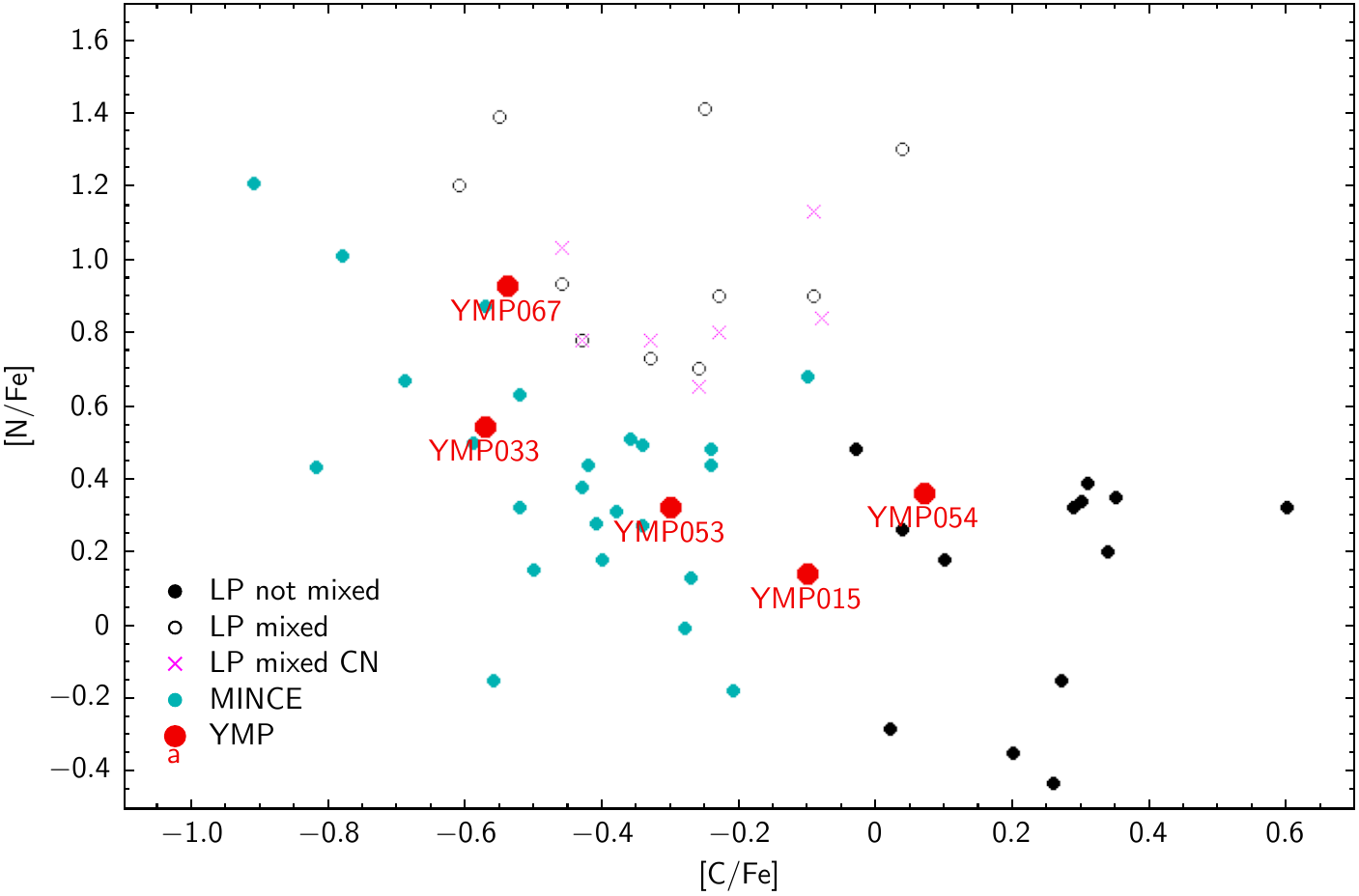}}
    \caption{The [N/Fe] vs. [C/Fe] diagram for our program stars. As reference samples we took
    the Large Programme (LP) stars of \citet{spite2005}, mixed (open black symbols)
    and not mixed (filled black symbols) and also the N abundances derived from the 388\,nm
    CN band (violet $\times$ symbols), and the MINCE collaboration giants (Lombardo et al. these proceedings, cyan symbols).}
    \label{fig:cfe_nfe}
\end{figure}

%In Fig.\,\ref{fig:agemass} we report the mass vs. age diagram for our stars.
From the kinematics our targets appear to belong to the halo, except, YMP015 that is a thick disc stars.
They all appear to be too young with respect to their kinematical classification (Bonifacio et al. in prep).
%Even YMP015, that is a thick disc star should have an age older than 8\,Gyr \citep{haywood2016};
%the halo stars should  have ages older than 10\,Gyr, compatible with the Galactic Globular Clusters. (GGC) \citep[see e.g.][]{gratton2003}.
They could be either truly young stars or evolved blue straggler stars (BSS).

In this contribution we investigate the CNO abundances in these stars to see if they can help
to discriminate between BSS and young stars.
\citet{ferraro2006} found among the BSS in the Galactic Globular Cluster  (GGC) 47\,Tuc a group of stars that are
strongly depleted in C and O and their are shown in our Fig.\,\ref{fig:cfe_ofe}. They interpret these stars as BSS formed
through mass transfer, since an extra mixing is expected to occur in this case
\citep{sarna1996}. The other BSS that are not strongly depleted
in C and O are interpreted as having been formed through collisions, since in this
process no extensive mixing is expected \citep{lombardi1995}.
In GGC the majority of BSS is expected to be formed
through collisions and, for example, C and O depleted BSS have been found
in the M\,30 \citep{lovisim30}, but not in NGC\,6752 \citep{lovisi6752}.
Among field stars, however we expect the majority of BSS to be formed by mass transfer \citep{PS2000} 
because stellar collisions are exceedingly rare.
If our stars were BSS  we should expect them to be from mass tranfer, so heavily depleted in C and O, which is clearly
not the case, as shown in Fig.\,\ref{fig:cfe_ofe}.
One complication is that our stars are evolved and \citet{ferraro2006} suggested that these depletions
may be temporary and as the BSS evolves the mixing restores the original abundances,
prior to the mass transfer. \citet{ferraro2006}, however, do no corroborate this suggestion
with any theoretical or observational evidence. In our case we note that the mass transferred
must be very large, in the range 0.2 to 0.5 $M_\odot$, assuming the accreting star has a mass
of 0.8\,$M_\odot$, even larger if its mass is smaller. Therefore we are not talking about a thin
layer that only pollutes the stellar photosphere.

Since we have the abundances of C, N and O, we can try to assess the mixing properties of our stars.
In Fig.\,\ref{fig:cfe_nfe} we plot the [N/Fe] vs. [C/Fe] diagram for our program stars with two comparison samples,
the MINCE sample (Lombardo et al. these proceedings) and the ESO Large Programme ``First Stars'' \citep[][hereafter LP]{spite2005}.
The MINCE sample has metallicities similar to our sample, but their nitrogen abundances were derived from the 
NH $A^3\Pi_i-X^3\Sigma^-$ band at 336\,nm. The LP is typically more metal-poor than our sample, all their N abundances
are also derived from the same UV NH band (black symbols), however for a subset of stars they also determined N abundances
from the 388\,nm band, like us.  Among the LP stars one sees discrepancies between the NH and CN bands, 
in both directions. From this plot it is clear that
YMP015, YMP033, YMP053 and YMP067 qualify as mixed stars, while YMP054 is not mixed. The comparison
with the MINCE sample, that occupies the same region in the diagram, suggests that
probably the abundances derived from NH and CN are consistent. Any discrepancies should not change
the classification of the star as mixed or not mixed. Our interpretation is that
the carbon deficiencies and nitrogen enhancement in these stars are compatible with
their evolutionary status and can be explained by mixing.

Our conclusion is that although we cannot exclude that these stars are evolved BSS, the abundances of CNO
point towards their being truly young. It is clear that the sample may contain both kinds of stars.
Wider searches are needed, especially aimed at finding more of these metal poor apparently young
stars and especially of higher masses. If a significant number stars with masses exceeding
1.6\,$M_\odot$ were selected, then it would be very difficult to explain them as evolved BSS.

%\section*{Acknowledgements}
%  We appreciate 
  
\bibliographystyle{aa}
\bibliography{ymp_cno}

\end{document}